\documentclass[aip,graphicx, reprint]{revtex4-1}
\usepackage{xcolor}
\usepackage{graphicx}
\usepackage{subfigure}
\usepackage{placeins}
\usepackage{amsmath}
\usepackage{float}
\usepackage[normalem]{ulem}

\draft 

\begin{document}

\title{Large critical current density Josephson $\pi$ junctions with PdNi barriers}

\author{Arjun Sapkota}
\affiliation{Materials Science, Engineering and Commercialization Program, Texas State University, San Marcos, Texas 78666, USA}

\author{Pukar Sedai}
\affiliation{Materials Science, Engineering and Commercialization Program, Texas State University, San Marcos, Texas 78666, USA}

\author{Robert M. Klaes}
\affiliation{Department of Physics and Astronomy, Michigan State University, East Lansing, Michigan 48824, USA}

\author{Reza Loloee}
\affiliation{Department of Physics and Astronomy, Michigan State University, East Lansing, Michigan 48824, USA}

\author{Norman O. Birge}
\affiliation{Department of Physics and Astronomy, Michigan State University, East Lansing, Michigan 48824, USA}

\author{Nathan Satchell}
 \email{satchell@txstate.edu}
 \affiliation{Materials Science, Engineering and Commercialization Program, Texas State University, San Marcos, Texas 78666, USA}
\affiliation{Department of Physics, Texas State University, San Marcos, Texas 78666, USA}

\date{\today}

\begin{abstract}
We report large $\pi$-state critical current densities, $J_c(\pi)$, in Nb/Pd$_{89}$Ni$_{11}$/Nb Josephson junctions at Pd$_{89}$Ni$_{11}$ thicknesses near the first $\pi$-state. We observe oscillations in the critical current with ferromagnetic barrier thickness consistent with a $0$--$\pi$ transition. For a junction with a 9.4~nm Pd$_{89}$Ni$_{11}$ barrier, we obtain $J_c(\pi) = 410~\mathrm{kA/cm^{2}}$ at 4.2~K, exceeding values reported in prior PdNi-based studies. Magnetization measurements on continuous films, together with coercivity tests on patterned arrays, confirm that Pd$_{89}$Ni$_{11}$ exhibits perpendicular magnetic anisotropy, enabling zero-field operation without magnetic initialization. The combination of large $J_c(\pi)$ and intrinsic anisotropy establishes Pd$_{89}$Ni$_{11}$ as a promising barrier material for passive $\pi$-shifters in superconducting digital logic and qubit architectures.

\end{abstract}

\pacs{}

\maketitle 

Josephson junctions with ferromagnetic barriers have attracted significant interest over the past two decades, particularly in the development of $\pi$-junctions.\cite{veretennikov_2000,Ryazanov_PRL_2001,Kontos_PRL_2002} The proposed applications of $\pi$-junctions include as components in highly energy efficient superconducting computing schemes and in certain qubit realizations.\cite{cai_2023, likharev_2012, soloviev2017beyond, birge_2024} The most promising routes towards utilizing $\pi$-junctions in these applications are proposals to incorporate $\pi$-junctions as passive phase-shifters, where the $\pi$-junctions are placed into superconducting loops to replace an applied flux bias.\cite{terzioglu_1998,ustinov_2003,ioffe_1999,blatter_2001,ortlepp_2006,mielke_2008,khabipov_2010,wetzstein_2011,kamiya_2018,yamanashi_2018,hasegawa_2019,arai_2019,yamanashi_2019,yamashita_2021,takeshita_2021,li_2021,li_2023,li_2023a,tanemura_2023,soloviev_2021,soloviev_2022,maksimovskaya_2022,khismatullin_2023,kato_2007,feofanov_2010,feofanov_2010,shcherbakova_2015,Kim_2024}

Successful implementation of the proposed applications requires $\pi$-junctions to have large critical currents, (greater than that of the insulating junctions in the loop), a maximum critical current achieved at or near zero applied magnetic field without magnetic initialization, and, in the case of qubits, for the junction dynamics to be underdamped.\cite{kato_2007} Achieving large critical current density, $J_c$, values in magnetic $\pi$-junctions is challenging due to the typically short ferromagnetic coherence length, but it can be mitigated in certain materials, such as Ni and some Ni alloys.\cite{Baek_PRApp_2017, birge_2024} The applied magnetic field requirements can be satisfied using ferromagnetic materials with perpendicular magnetic anisotropy (PMA).\cite{Satchell_SciRep_2021} For qubits, the requirement to be underdamped cannot be met in all-metal junctions. However, ferromagnetic insulators or metallic junctions that incorporate an additional insulating layer in series with a metallic ferromagnet hold promise.\cite{caruso_2019, Kontos_PRL_2002, Weides_APL_2006, Ahmad_2024}

Experimental realizations of $\pi$-junctions have now been demonstrated in many magnetic materials; see Ref.~\onlinecite{birge_2024} for a recent review. Among these systems, PdNi has emerged as a promising candidate barrier material.\cite{Kontos_PRL_2002,Khaire_PRB_2009,Pham_2022,li_2023a} It exhibits intrinsic perpendicular magnetic anisotropy (PMA), and prior work by Pham \textit{et al.} showed that all-metallic Pd$_{89}$Ni$_{11}$ junctions with NbN electrodes can reach $J_c \approx 70~\mathrm{kA/cm^2}$ in the first $\pi$-state at $4.2$~K.\cite{Pham_2022}

Here, we fabricate and measure Nb-based Josephson junctions with Pd$_{89}$Ni$_{11}$ (hereafter PdNi) barriers, focusing on barrier thicknesses near the first $\pi$-state. Our primary objective is to experimentally access this regime and quantify the resulting $\pi$-state critical current density, $J_c(\pi)$. We find that Nb/PdNi/Nb junctions exhibit exceptionally large $J_c(\pi)$ values: for a $9.4$~nm PdNi barrier, we obtain $J_c(\pi)= 410~\mathrm{kA/cm^2}$ at $4.2$~K for a $\sim 7~\mu\mathrm{m}^2$ circular device, exceeding previously reported PdNi-based junctions.\cite{Pham_2022} Thickness-dependent measurements within the first $\pi$-state regime indicate that this performance is robust against modest variations in barrier thickness. Magnetometry further confirms that PdNi exhibits intrinsic perpendicular magnetic anisotropy, enabling zero-field operation without magnetic initialization. The combination of large $J_c(\pi)$ and intrinsic anisotropy directly addresses key device-level constraints for passive $\pi$-shifters in superconducting digital logic and qubit architectures.

The films are deposited onto 0.5 mm thick Si substrates with 100 nm thermal oxide layer using dc sputtering from targets with a typical purity of 99.99\%. Deposition of the bottom 60 nm Nb electrode, PdNi barrier layer of varying thickness, and 10 nm Pd capping layer are performed without breaking vacuum in a custom built sputtering system with base pressure of $5 \times 10^{-8}~\text{Torr}$. The PdNi alloy is deposited by co-sputtering from separate Pd and Ni targets. Deposition rates and alloy composition are calibrated using an \textit{in situ} crystal film thickness monitor. The films are patterned to circular Josephson junctions of diameter $3~\mu\text{m}$ using standard photolithography and ion milling methods, described in a previous work.\cite{Wang_2012} In the final stage of fabrication, the samples are loaded into a dc sputtering system with base pressure $2 \times 10^{-8}~\text{Torr}$. The 10 nm Pd capping layer is partially ion milled \textit{in situ}, thus recovering a very clean interface before depositing the top 150 nm Nb electrode. 

To characterize the magnetic properties of PdNi, we first fabricated a 35~nm continuous film. 
To assess whether micro-patterning alters the coercive field, we fabricated a second sample consisting of a circular dot array with lateral dimensions matching those of the Josephson devices. 
The pattern was defined by standard lift-off photolithography and consists of $3~\mu$m circular elements on a $9~\mu$m center-to-center pitch. 
PdNi was sputtered to a thickness of 35~nm into both the patterned substrate and a companion unpatterned substrate used as a continuous-film reference. To examine the effect of reduced thickness on the magnetic properties, an additional 10~nm continuous PdNi film was fabricated.
Magnetization loops were measured using a Quantum Design Physical Property Measurement System (PPMS) with the Vibrating Sample Magnetometer (VSM) option.
Electrical transport measurements of the Josephson junctions were performed in the same PPMS using the Horizontal Rotator option and a Lake Shore M81 synchronous source-measure system. 
The magnetic field during transport measurements was applied parallel to the sample plane.

\begin{figure}
    \centering
    {\includegraphics[width=\linewidth]{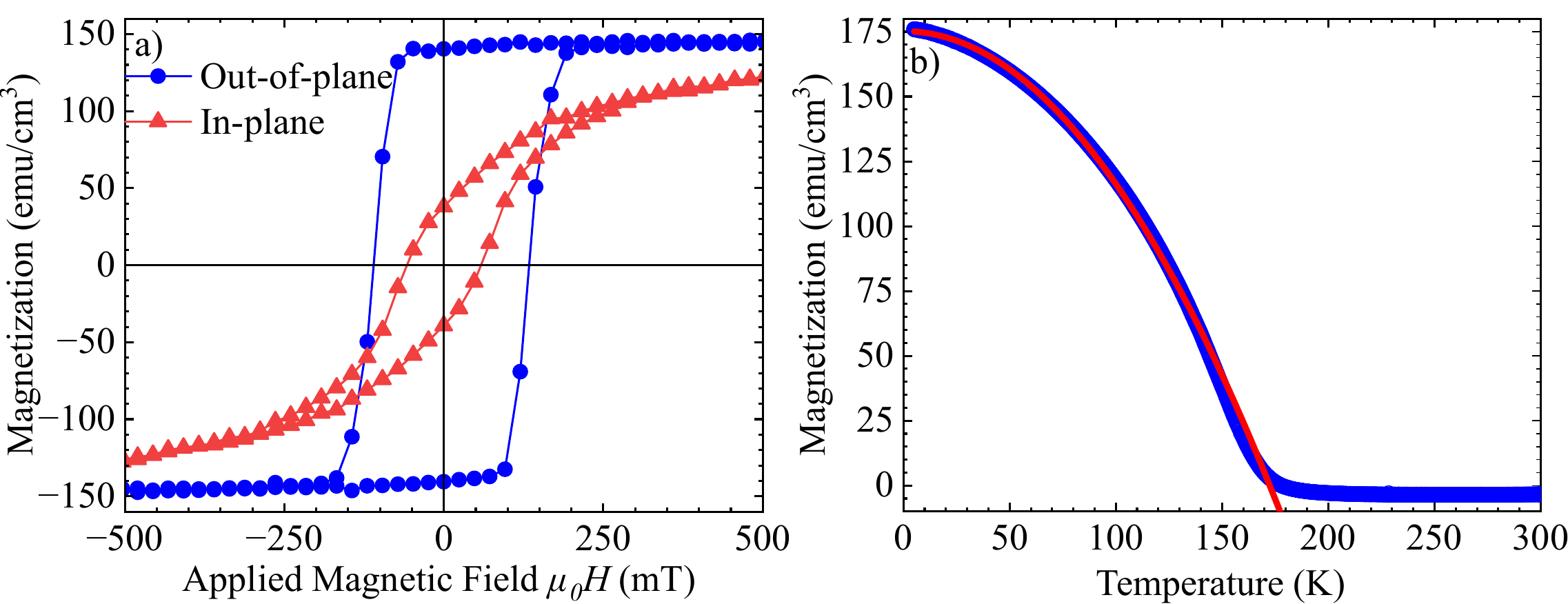}}
    \caption{Magnetic characterization of continuous 35-nm-thick single-layer $\text{Pd}_{89}\text{Ni}_{11}$ film. (a) Magnetic field-dependent magnetization at 5 K with the applied field out-of-plane and in-plane. (b) Temperature-dependent magnetization for an out-of-plane applied field of 1~T. The data are fitted to Eq.~\ref{eq:MT}, and the Curie temperature is approximately 170 K. Magnetization values are calculated from the nominal film thickness, measured total magnetic moment, and the areas of the samples. The measurements were performed using vibrating sample magnetometry. The uncertainty in each point is dominated by the area measurements, with an uncertainty of less than 10\%. Lines connecting the data points are intended as guides for the eye.}
    \label{fig:1}
\end{figure}

Fig. \ref{fig:1} (a) shows the magnetization as a function of the applied magnetic field measured both in-plane and out-of-plane for a continuous 35 nm thick single layer PdNi thin film at 5~K. The total magnetic moment of the sample is measured; then we subtract the linear background due to the substrate and estimate the volume of the PdNi film from the nominal film thickness and measured sample area to calculate the magnetization of the PdNi. The film shows PMA, as indicated by the higher squareness ratio of the field out-of-plane loop compared to the in-plane field orientation. The saturation magnetization of the PdNi film at 500~mT out-of-plane applied field is determined to be $145 \pm 15~\text{emu/cm}^3$.

\begin{figure}
    \centering
    \includegraphics[width=0.95\linewidth]{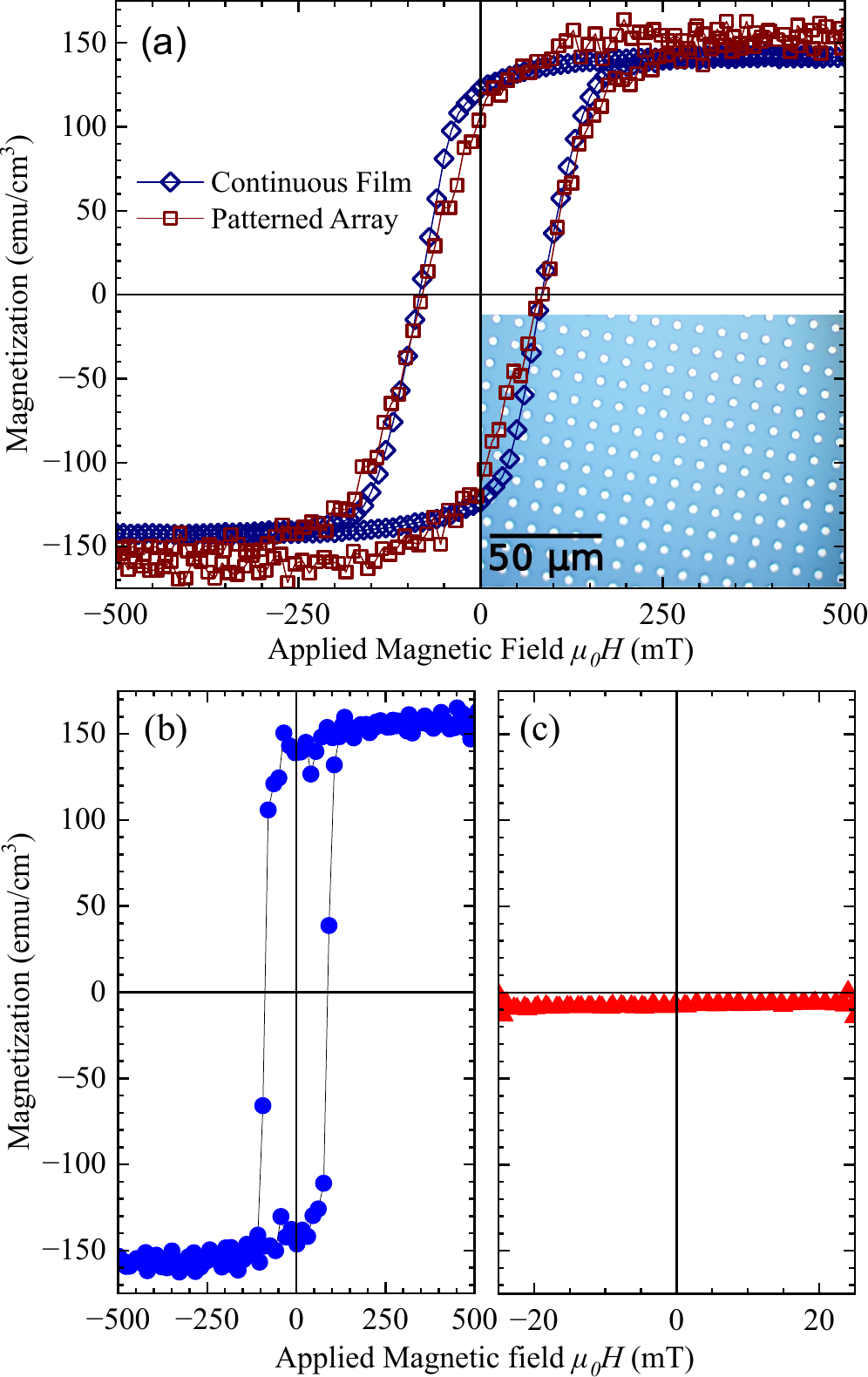}
    \caption{
(a) Out-of-plane magnetization at 5~K for two Pd$_{89}$Ni$_{11}$ films fabricated in the same deposition run: 
a 35~nm continuous film and a patterned array of circular elements with measured diameter 
$\sim 3.5~\mu$m on a $9~\mu$m pitch. The inset shows an optical micrograph of the array; image analysis yields an areal fill factor of $\sim 12\%$. 
(b) Out-of-plane magnetization loop at 5~K for an additional 10~nm continuous Pd$_{89}$Ni$_{11}$ film. 
(c) In-plane minor hysteresis loop for the patterned array measured within $\pm 25$~mT, corresponding to the field range used in the Josephson junction measurements.
Magnetization values were calculated from the total moment, nominal film thickness, and sample area. Lines connecting data points are guides to the eye.
}
    \label{fig:2}
\end{figure}

Fig. \ref{fig:1} (b) shows the temperature dependence of magnetization, $M(T)$, for an out-of-plane applied magnetic field of 1~T. $M(T)$ can be described by the Bloch law\cite{bloch_1930}
\begin{equation}
M(T) = M(0) \left[ 1 - \left( \frac{T}{T_{\text{Curie}}} \right)^\alpha \right],
\label{eq:MT}
\end{equation}
where $M(0)$, $T_{\text{Curie}}$, and $\alpha$ represent the magnetization at 0~K, the Curie temperature, and Bloch’s exponent, respectively. The best fit shown in Fig. \ref{fig:2} suggests that $\alpha$ is 2.1 and the Curie temperature is approximately 170~K, consistent with other values in the literature for thin PdNi films of similar Ni concentration.\cite{Khaire_PRB_2009,Pham_2022}

To determine whether patterning PdNi at junction-scale lateral dimensions modifies its magnetic response, we fabricated two additional 35~nm thick PdNi films in the same deposition run: a continuous film and an array of circular elements. Optical microscopy of the patterned sample (Fig.~\ref{fig:2}, inset), combined with threshold-based image analysis, yields an average element diameter of $3.5~\mu$m with a center-to-center spacing of $9~\mu$m, corresponding to an areal fill factor of $\sim 12\%$. Figure~\ref{fig:2} shows the out-of-plane $M(H)$ loops measured at 5~K for the two samples. The coercive field, remanent magnetization, and perpendicular easy axis of the patterned array are indistinguishable from those of the continuous film within experimental uncertainty, indicating that patterning at $\sim 3.5~\mu$m lateral dimensions does not alter the intrinsic magnetic response of PdNi. No additional features consistent with collective switching or strong dipolar coupling between elements are observed. A systematic investigation of spacing-dependent dipolar interactions was not undertaken here. Because our junction dimensions remain in the micrometer regime, these conclusions should not be extrapolated to substantially smaller, nanometer-scale magnetic elements.\cite{cowburn2000property, RevModPhys.78.1}

We further examine the thickness dependence of PdNi magnetization by measuring an additional 10~nm continuous film, Fig.~\ref{fig:2} (b). The 10~nm film exhibits a square out-of-plane hysteresis loop at 5~K, consistent with PMA. The saturation magnetization is comparable to that of the 35~nm films and patterned array, indicating that reduction of the thickness to 10~nm does not significantly alter the overall magnetic behavior. In addition, Fig.~\ref{fig:2} (c) shows an in-plane minor hysteresis loop of the patterned array measured within $\pm 25$~mT, corresponding to the magnetic field range used for Josephson junction measurements.

\begin{figure}
    \centering
    \includegraphics[width=\linewidth]{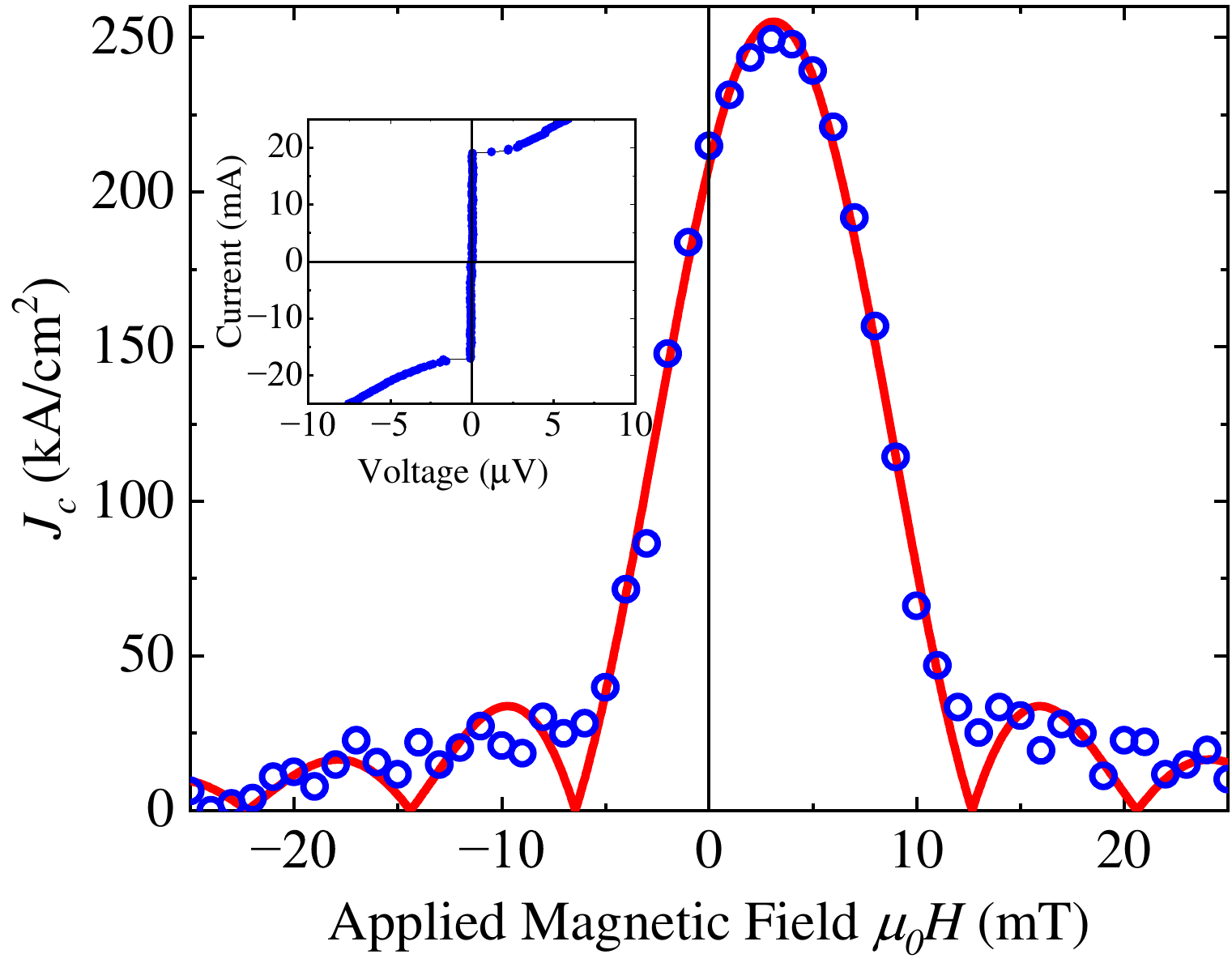}
    \caption{Critical current density, $J_c$, versus applied magnetic field for the  $ \text{Nb} / \text{Pd}_{89}\text{Ni}_{11} / \text{Nb} $ Josephson junction with $\text{Pd}_{89}\text{Ni}_{11}$ thickness of 9.4 nm at 5.75 K. The $J_c$ is calculated from the nominal size of the 3~\textmu m diameter circular junction and measurements of the $I-V$ characteristic at each field. The solid red line is a fit to the data of Eqs. \ref{eq:Ic} and \ref{eq:phii}. Inset: The $I-V$ characteristic of the same 3~$\mu$m  circular  junction at zero applied magnetic field. The critical current, $I_c$, is determined from these data by fitting to Eq. \ref{eq:VI}.}
    \label{fig:3}
\end{figure}

Fig. \ref{fig:3} shows electrical transport characterization of the Nb/PdNi/Nb Josephson junction with a PdNi thickness of 9.4~nm. At each field, we measure the current-voltage ($I-V$) characteristic. An example for $I-V$ at zero applied field is shown in the inset of Fig. \ref{fig:3}. The $I-V$ curves indicate our junctions are overdamped, and can therefore be described by\cite{barone1982physics} 
\begin{equation}
V(I) = \frac{I}{|I|} R_N \sqrt{I^2 - I_c^2} \quad \text{for } I \geq I_c,
\label{eq:VI}
\end{equation}
where $I_c$ is the critical current through the junction and $R_N$ is the normal state resistance across the junction. 

For the rest of this letter, we will report the nominal $J_c$ values of our junctions. For analysis and comparison with other literature, it can be helpful to convert between the values of $J_c$ and the product $I_c R_N$, which can be achieved by applying the relation $J_c = I_c R_N/A R_N$, where $A$ is the area of our 3~$\mu$m circular junctions and the average product $A R_N$ is $4.4 \pm 0.3\ f\Omega$m$^2$.

When a magnetic field is applied perpendicular to the current across the junction, the critical current at each applied field shows a “Fraunhofer” interference pattern, as shown in Fig. \ref{fig:3}. Since the junctions are circular, the dependence of $J_c$ with applied magnetic field can be described by the Airy function\cite{barone1982physics}

\begin{equation}
J_c ({\Phi}) = J_{c}(\text{max}) \left| \frac{2J_1 \left(\pi \frac{\Phi}{\Phi_0} \right)}{\pi \frac{\Phi}{\Phi_0}} \right|,
\label{eq:Ic}
\end{equation}
where $J_{c}(\text{max})$ is the maximum critical current through junction, $J_1$ is a Bessel function of the first kind, $\Phi_0$ is the flux quantum ($h/2e$) and $\Phi$ is the flux through the junction, given by\cite{barone1982physics}
\begin{equation}
\begin{aligned}
\Phi &= \mu_0 (H_{\text{app}} - H_{\text{shift}}) w \Bigg[
\lambda_L^{\text{bottom}} \tanh \left( \frac{d_S^{\text{bottom}}}{2\lambda_L^{\text{bottom}}} \right) \\
&\quad + \lambda_L^{\text{top}} \tanh \left( \frac{d_S^{\text{top}}}{2\lambda_L^{\text{top}}} \right) + d_F
\Bigg],
\label{eq:phii}
\end{aligned}
\end{equation}
where $w$ is the width of the junction, $H_{\text{app}}$ is the applied field,  $H_{\text{shift}}$ is amount of field the $J_{c}(\text{max})$ is shifted from $H=0$, $d_S$ is the thickness of the superconducting electrodes, and $d_F$ is the thickness of PdNi ranging from 6.8 to 14.1~nm in this study. Since both electrodes are Nb, we set $\lambda_L^{\text{bottom}} = \lambda_L^{\text{top}}$ = 100~nm.\cite{quarterman2020distortions} Small nonzero values of $H_{\text{shift}}$ are found to be independent of PdNi thickness. Measurements of the patterned PdNi array under the same in-plane field range used for the junction measurements, Fig.~\ref{fig:2} (c), show no significant hysteresis or remanent magnetization.
Taken together, these observations indicate that the small, thickness-independent $H_{\text{shift}}$ most likely originates from trapped flux in the superconducting solenoid used to apply the magnetic field, rather than from in-plane magnetization of the ferromagnetic barrier.

We follow the same measurement and analysis procedure for each junction in our study. Experimentally, we found that the critical currents of our junctions are very high. To help mitigate the risk of Joule heating from large applied currents, we performed the Fraunhofer measurements at 5.75~K, where the increased temperature reduces the critical current. We report on the temperature dependence of $J_c$ later in this letter.

\begin{figure}
    \centering
    \includegraphics[width=\linewidth]{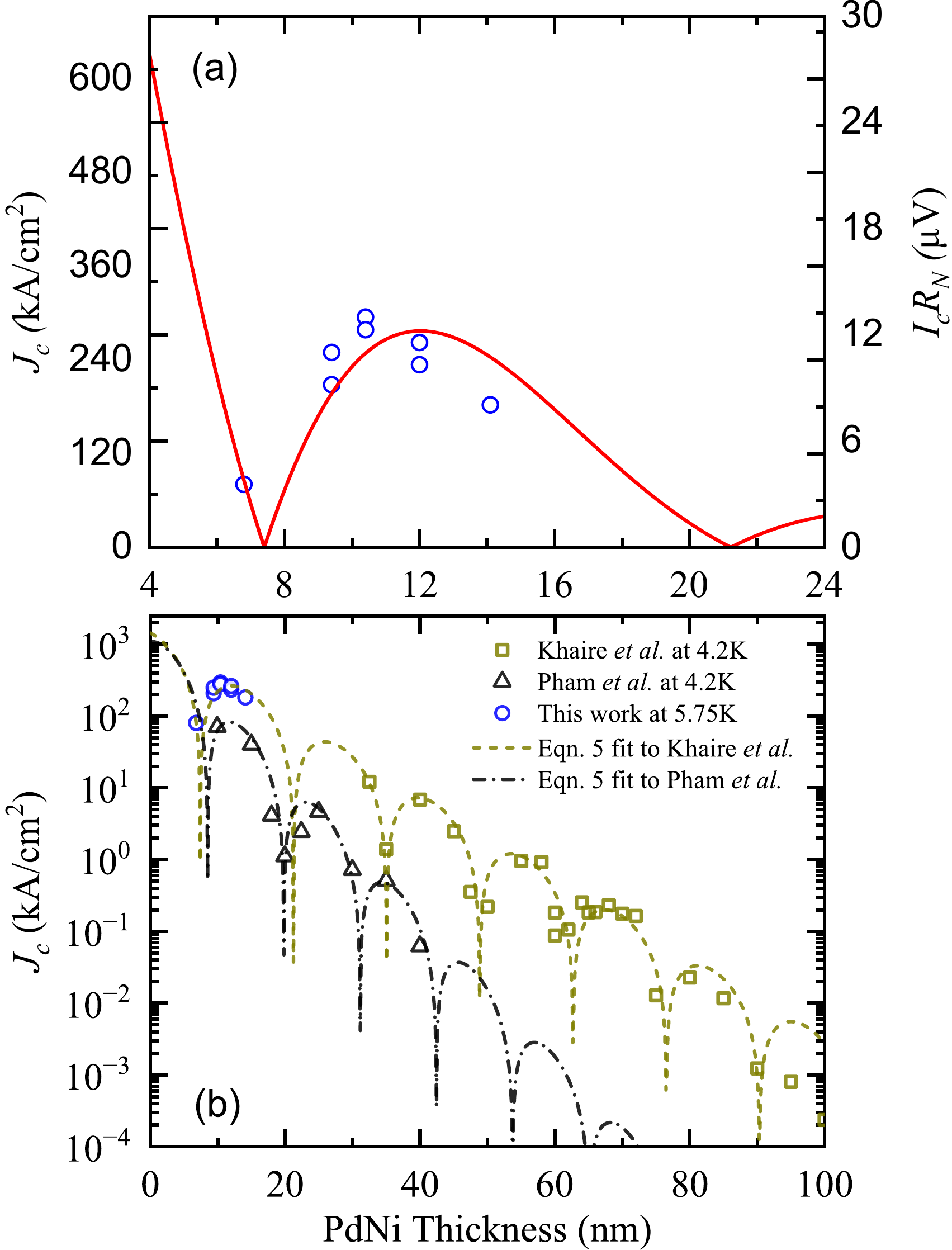}
    \caption{(a) The product of the maximum critical current and normal state resistance, $I_cR_N$ (right axis) plotted with the critical current density, $J_c$ (left axis) of the $ \text{Nb} / \text{Pd}_{89}\text{Ni}_{11} / \text{Nb} $ Josephson junctions versus nominal $\text{Pd}_{89}\text{Ni}_{11}$ thickness at 5.75~K. The data are described by Eq. \ref{eq:ic_rn} assuming the extrapolated fitting parameters of  Khaire \textit{et al.}~\cite{Khaire_PRB_2009} Each data point represents one Josephson junction, and the uncertainty in determining $I_c R_N$ and $J_c$ is smaller than the data points and (b) comparison of this work (blue circle) at 5.75~K with the work of Khaire \textit{et al.}~\cite{Khaire_PRB_2009}(green square) and Pham \textit{et al.}~\cite{Pham_2022} (black triangle) at 4.2~K.}
    \label{fig:4}
\end{figure}

Figure~\ref{fig:4} (a) shows the $I_cR_N$ and $J_c$ for each junction at 5.75~K, plotted together to allow the reader to compare the values with previous literature. An oscillation in $J_c$ with increasing thickness of PdNi would indicate a transition between the 0- and $\pi$-states. In this study, we measure only a limited range of thicknesses, chosen based on prior work to be close to the expected first $\pi$-state. In the intermediate limit, the decay and oscillation of $J_c$ (or $I_cR_N$) can be described by:\cite{bergeret_2001a}
\begin{equation}
J_c = J_c(0) \exp\left(\frac{-d_F}{\xi_{F1}}\right) \left| \sin\left( \frac{d_F - d_{\text{0-}\pi}}{\xi_{F2}} \right) \right|,
\label{eq:ic_rn}
\end{equation}
where $J_c(0)$, $d_{0-\pi}$, $\xi_{F1}$, and $\xi_{F2}$ are the fictitious zero-thickness fit parameter, the thickness of the first 0--$\pi$ transition, and the length scales governing the decay and oscillation of $J_c$, respectively. Figure~\ref{fig:4} (b) shows the comparison of our work with the work of Khaire \textit{et al.}~\cite{Khaire_PRB_2009} and Pham \textit{et al.}~\cite{Pham_2022} and the corresponding data are fitted with Eq.~\ref{eq:ic_rn}.   Revisiting prior work on PdNi by Khaire \textit{et al.}, the extrapolated fitting parameters are $J_c(0) = 1500~\text{kA/cm}^2$, $d_{0-\pi} = 7.4$~nm, $\xi_{F1} = 7.7$~nm, and $\xi_{F2} = 4.4$~nm.\cite{Khaire_PRB_2009} Fixing $d_{0-\pi}$, $\xi_{F1}$, and $\xi_{F2}$ values in Eq.~\ref{eq:ic_rn} to those of Khaire \textit{et al.} and allowing only $J_c(0)$ to be a free parameter in the fit yields a fortuitous agreement with our experimental data for $J_c(0) = 1500 \pm 100\,\text{kA/cm}^2$. We note, however, several important caveats: $J_c(0)$ is a phenomenological fitting parameter provided only to reproduce the presented fits; it is temperature dependent, and our measurements were performed at 5.75~K, compared to 4.2~K in the work of Khaire \textit{et al.}, making it reasonable to treat this parameter as free; our samples were fabricated using a different sputtering system and by cosputtering rather than by employing a stoichiometric target, although the similarity in Curie temperatures between our samples and those of Khaire \textit{et al.} suggests comparable magnetic behavior. Finally, although a closer fit of Eq.~\ref{eq:ic_rn} to our experimental data could be achieved by allowing all parameters to vary freely, we refrain from extracting new fitting parameters, as the narrow range of thicknesses studied does not provide sufficient confidence in their reliability. Nevertheless, we find it remarkable that the fit parameters extracted by Khaire \textit{et al.} from junctions with PdNi thicknesses ranging from 30 to 100~nm matched exactly with our first $\pi$-state data from junctions with PdNi thicknesses between 7 and 14~nm, as shown in Fig.~\ref{fig:4} (b). The only difference lies in the measurement temperature: we measured at 5.75~K, whereas Khaire \textit{et al.} measured at 4.2~K.

\begin{figure}
    \centering
    \includegraphics[width=\linewidth]{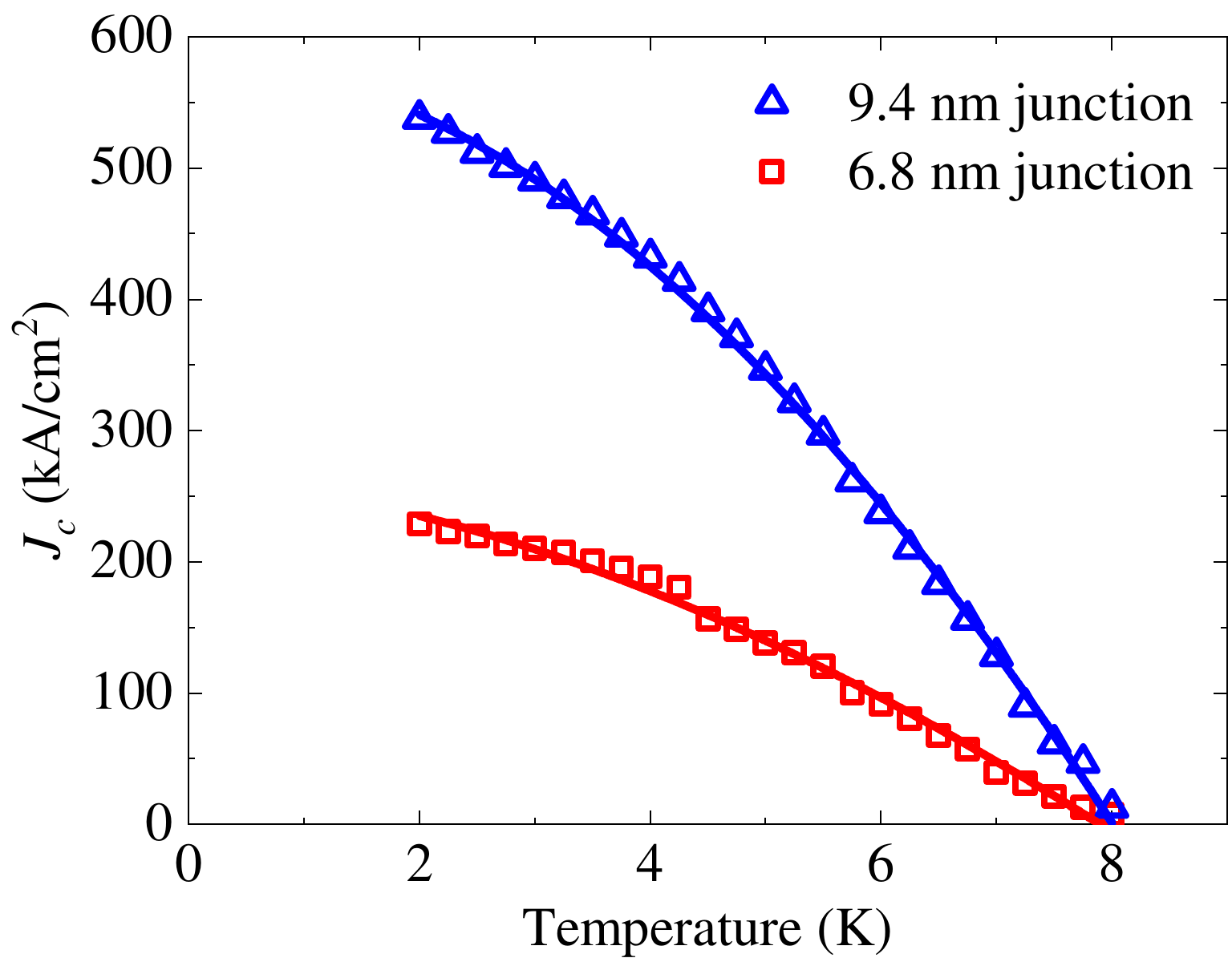}
    \caption{Critical current density, $J_c$, versus temperature at zero magnetic field and without any magnetic initialization for the $ \text{Nb} / \text{Pd}_{89}\text{Ni}_{11} / \text{Nb} $ Josephson junctions with $ \text{Pd}_{89}\text{Ni}_{11} $ thicknesses of 6.8~nm and 9.4~nm. The critical temperature is approximately 8~K, and the trend of $ J_c (T) $ is well described by the phenomenological expression, Eq.~\ref{eq:J_c(T)}. The data points at each temperature are averaged from three $ I-V $ curve measurements, and the statistical uncertainty in $J_c$ is smaller than the data points.
}
    \label{fig:5}
\end{figure}

Fig. \ref{fig:5} shows $J_c$ for junctions with PdNi 6.8~nm and 9.4~nm at varying temperatures. Phenomenologically, $J_c(T)$ can be described as,
\begin{equation}
J_c(T) = J_c(\text{0~K}) \left[ 1 - \left( \frac{T}{T_{\text{c}}} \right)^n \right],
\label{eq:J_c(T)}
\end{equation} 
where $J_c(\text{0~K})$ is the extrapolated critical current density at 0~K, $T_{\text{c}}$ is the critical temperature, and $n$ is an empirically determined exponent. $J_c$ is largest at the lowest temperatures and decreases with increasing temperature, approaching nearly zero at 8~K, close to the critical temperature of our Nb electrode. The best fits to our data shown in Fig~\ref{fig:5} are obtained using values for $J_c(\text{0~K})$, $T_{\text{c}}$, and $n$ of $584\pm8$~kA/cm$^2$, $8.01\pm0.02$~K, $1.88\pm0.05$, respectively for the 9.4~nm junction and $261\pm8$~kA/cm$^2$, $7.90\pm0.06$~K, $1.7\pm 0.1$ for the 6.8~nm junction.

The difference in $J_c$ between these two junctions can be understood by studying the 0-$\pi$ oscillations in Fig. \ref{fig:4}(a). The 9.4~nm junction is near the peak of the $\pi$-state, where $J_c$ is expected to be large, whereas the 6.8~nm junction is close to the 0-$\pi$ transition, where $J_c$ is expected to be small.

We next consider the role of junction size. Josephson junctions are commonly classified as
``large'' when their lateral dimensions exceed the Josephson penetration depth,
$\lambda_J=\sqrt{\hbar/[2e\mu_0(2\lambda_L+d_F)J_c]}$.\cite{barone1982physics}
Using parameters appropriate for our Nb/PdNi/Nb junctions
($\lambda_L=100$~nm, $d_F=9.4$~nm, and $J_c(\pi)=410~\mathrm{kA/cm^2}$ at 4.2~K),
we obtain $\lambda_J=0.55~\mu$m. Since the lateral dimension of our circular junctions
(3~$\mu$m diameter) exceeds $\lambda_J$, the devices fall within the large-junction regime. In this limit, self-field effects and nonuniform current distributions can reduce the measured
critical current relative to the intrinsic value. Consequently, the experimentally determined
$J_c(\pi)$ values reported here likely represent a lower bound on the intrinsic $\pi$-state
critical current density, and smaller PdNi junctions fabricated at the same barrier thickness
may support even larger $J_c(\pi)$.

To place our results in context, we first compare our junctions to prior studies of PdNi barriers, and then to ferromagnetic $\pi$-junctions more broadly. In the context of PdNi, we have noted a fortuitous agreement with the work of Khaire \textit{et al.},\cite{Khaire_PRB_2009} highlighted in Fig.~\ref{fig:4}(b). Because $J_c(0)$ is a fictitious fitting parameter, we compare experiments using the directly measurable quantity $J_c(\pi)$, the maximum critical current density attained in the first $\pi$-state. Khaire \textit{et al.} did not experimentally access the first $\pi$-state, but extrapolating their fit in Fig.~\ref{fig:4}(b) to $d_F\approx 12$~nm, suggests $J_c(\pi)\approx 260~\mathrm{kA/cm^2}$ at $4.2$~K. Their measurements were performed on Nb-based circular junctions of $10$-$80~\mu$m diameter ($\sim 78$–$5026~\mu\mathrm{m}^2$). Pham \textit{et al.}~\cite{Pham_2022} report $J_c(\pi)\approx 70~\mathrm{kA/cm^2}$ at $4.2$~K. Their measurements were performed on NbN-based $10~\mu\mathrm{m}$ square junctions ($100~\mu\mathrm{m}^2$). In the present work, the $d_F = 9.4$~nm device exhibits $J_c(\pi)=410~\mathrm{kA/cm^2}$ at $4.2$~K, i.e., substantially larger than in either prior study. We use Nb-based circular junctions of $3~\mu$m-diameter ($7.1~\mu\mathrm{m}^2$).

Pham \textit{et al.} extract an effective interface transparency, $\mathcal{T}$, and parameter, $\gamma_B$, using a linearized Usadel analysis in the diffusive limit.\cite{Pham_2022} Applying the same methodology to our devices enables a direct comparison using identical assumptions. Using our measured average $AR_N = 4.4 \pm 0.3~\mathrm{f\Omega\,m^2}$ and the PdNi resistivity $\rho_F = 68~\mathrm{n\Omega\,m}$ reported by Khaire \textit{et al.},\cite{Khaire_PRB_2009} we obtain $\gamma_B \approx 3.5$ and effective transparency $\mathcal{T} \approx 0.20$. Repeating the procedure using the parameters of Pham \textit{et al.}, including $\rho_F = 360~\mathrm{n\Omega\,m}$, yields $\gamma_B \approx 4.8$ and $\mathcal{T} \approx 0.12$. Within the same theoretical framework, the Nb/PdNi interfaces are therefore substantially more transparent than the NbN/PdNi interfaces. The higher interface transparency, together with the longer superconducting coherence length of Nb relative to NbN, provides a possible explanation for the larger $\pi$-state $J_c$ observed in our devices, independent of junction area or geometry.

Comparing our work to other ferromagnetic $\pi$-junctions, we note that the only published
systems that reach higher $J_c(\pi)$ values are those based on elemental ferromagnets such as
Ni.\cite{Robinson_PRL_2006, Robinson_PRB_2007, Piano_EPJB_2007, Robinson_APL_2009,
Baek_PRApp_2017, Dayton_IEEE_2017, Kapran_2021} Although such junctions can carry very
large supercurrents, they lack intrinsic perpendicular magnetic anisotropy and typically require
magnetic initialization in relatively large in-plane fields to produce reproducible Fraunhofer
patterns. In contrast, PdNi offers two intrinsic advantages. First, its PMA yields a centered and
robust Fraunhofer pattern in the virgin state. Second, its moderate exchange field produces a
significantly longer ferromagnetic coherence length than elemental ferromagnets, so the first
$\pi$-state occupies a broader thickness window, reducing the sensitivity of $J_c(\pi)$ to
nanometer-scale thickness variations.

In conclusion, we have fabricated and characterized Nb/PdNi/Nb Josephson junctions with PdNi barrier thicknesses ranging from 6.8 to 14.1~nm, spanning the regime where the first $\pi$-state is expected in this alloy. Magnetization measurements confirm that PdNi exhibits intrinsic perpendicular magnetic anisotropy, with a saturation magnetization of $145 \pm 15~\mathrm{emu/cm^3}$ and a Curie temperature of $\sim 170$~K. Additional patterned-array measurements show that this anisotropy is preserved at lateral dimensions comparable to our junctions. Our transport measurements reveal two central results. First, the thickness dependence of $J_c$ shows the onset of an oscillatory behavior consistent with a first $0$--$\pi$ transition, in quantitative agreement with the decay and oscillation scales reported by Khaire \textit{et al.} for thicker PdNi junctions. Second, junctions in the first $\pi$-state exhibit exceptionally large $\pi$-state critical current densities. For a 9.4~nm PdNi barrier, we obtain $J_c(\pi)= 410~\mathrm{kA/cm^2}$ at $4.2$~K and a zero-field value exceeding $550~\mathrm{kA/cm^2}$ at 2~K, without magnetic initialization. These values exceed those reported in previous PdNi-based studies and approach those found only in $\pi$-junctions based on elemental ferromagnets, while retaining the material advantages of PdNi: intrinsic PMA and a comparatively long ferromagnetic coherence length. The combination of large $J_c(\pi)$, robust zero-field operation, and tolerance to modest barrier-thickness variations identifies Pd$_{89}$Ni$_{11}$ as a compelling barrier material for passive $\pi$-shifters in superconducting digital logic and qubit architectures.

\begin{acknowledgments}
We acknowledge experimental assistance through the Analysis Research Service Center from Sam Cantrell and Casey Smith, fabrication assistance using the Keck Microfabrication Facility from Baokang Bi, and assistance with electrical transport measurements from Jonah Hartley and Kelsey Robbins.  We acknowledge support from new faculty startup funding made available by Texas State University.
\end{acknowledgments}

\section*{AUTHOR DECLARATIONS}

\subsection*{Conflict of Interest Statement}

The authors have no conflicts to disclose.

\subsection*{Author Contributions}

\textbf{Arjun Sapkota}: Software (lead), Validation (equal), Formal analysis (lead), Investigation (lead), Data Curation (lead), Writing – Original Draft (lead), Writing – Review \& Editing (equal), Visualization (lead). \textbf{Pukar Sedai}: Investigation (supporting), Writing – Review \& Editing (equal). \textbf{Robert M. Klaes}: Methodology (equal), Investigation (supporting), Writing – Review \& Editing (equal). \textbf{Reza Loloee}: Methodology (equal), Writing – Review \& Editing (equal). \textbf{Norman O. Birge}: Methodology (equal), Resources (equal), Writing – Review \& Editing (equal), Supervision (supporting). \textbf{Nathan Satchell}: Conceptualization (lead), Methodology (equal), Validation (equal), Investigation (lead), Resources (equal), Writing – Original Draft (supporting), Writing – Review \& Editing (equal), Supervision (lead), Project Administration (lead).

\section*{Data Availability Statement}

The data that support the findings of this study are openly available in the Texas Data Repository at \url{https://doi.org/10.18738/T8/XR2OF9}.

\bibliography{APLMatRev,APLMatRev_adds}

\end{document}